\documentclass[structabstract]{aa}  

\usepackage{graphicx}
\usepackage{txfonts}
\begin{document}

\title{High cadence spectropolarimetry of moving magnetic features \\ observed around a pore}

\author{Criscuoli~S. \inst{1}\thanks{currently at National Solar Observatory, Sacramento Peak, P.O. Box 62, Sunspot, NM 88349, USA}, Del~Moro~D.\inst{2}, Giannattasio~F. \inst{2},
Viticchi\'e~B.\inst{3,2}, Giorgi~F. \inst{1},\\ Ermolli~I. \inst{1}, Zuccarello~F. \inst{4},
Berrilli~F. \inst{2}}
   
\offprints{serena.criscuoli@oa-roma.inaf.it}
			 
\institute{INAF - Osservatorio Astronomico di Roma, Via Frascati 33, I-00040, Monte Porzio Catone, Italy  
\and Dipartimento di Fisica, Universit\`a degli Studi di Roma ``Tor Vergata'', Via della Ricerca Scientifica 1, I-00133 Roma, Italy   
\and ESA/ESTEC RSSD , Keplerlaan 1, 2200 AG Noordwijk, The Netherlands
\and Dipartimento di Fisica e Astronomia - Sezione Astrofisica, Universit\`a di Catania, Via S. Sofia 78, I-95123 Catania, Italy}     

\date{}

\abstract
	%CONTEXT
	{Moving magnetic features (MMFs) are small-size magnetic elements that
	are seen to stream out from sunspots, generally during their decay phase. Several observational results presented in the literature suggest them to be closely related to magnetic filaments that extend from the penumbra of the parent spot. 
	Nevertheless,  few observations of MMFs streaming out from spots without penumbra have been reported. The literature still lacks of
	analyses of the physical properties of these features.}
	% AIMS
	{We investigate physical
	properties of monopolar MMFs observed around a small
	pore that had developed  penumbra in the days preceding our observations and compare our results with those reported in the literature for features observed around sunspots.
	}
	%METHOD
	{We analyzed NOAA 11005 during its decay phase with data acquired
	at the Dunn Solar Telescope in the \ion{Fe}{i}~$617.3$~nm and the
	\ion{Ca}{ii}~$854.2$~nm spectral lines with IBIS, and in the G-band. The field of view showed monopolar 
	MMFs of both polarities streaming out from the leading negative polarity pore of the
	observed active region. Combining different analyses of the data, we
	investigated the temporal evolution of the relevant physical quantities
	associated with the MMFs as well as the photospheric and chromospheric signatures of
	these features.}
	%RESULTS
	{We show that the characteristics of the investigated MMFs agree with those reported in the literature for MMFs that stream out from spots with penumbrae. Moreover, observations of at least two of the observed features suggest them to be manifestations of emerging magnetic arches. 
}


\authorrunning{S. Criscuoli et al.}
\titlerunning{Spectropolarimetry of MMFs observed around a pore}
\maketitle

\section{Introduction}
\label{intro}
	Moving magnetic features (MMFs) are small-size magnetic elements seen to stream out from sunspots, especially during their decay phase. These features can appear as isolated features of polarity the same or opposite with respect to the spot they stream from (monopolar), or as couples of  features of opposite polarities (dipolar). Their horizontal velocities are usually below 1~km~s$^{-1}$ for  dipolar and monopolar features with same polarity as the spot, while higher velocities have been reported for monopolar features with opposite polarity with respect to the spot. 
	%These magnetic patches have been named Moving Magnetic Features (MMFs) by
	%\cite{harvey1973}.

	Recent observations have shown that MMFs form in the  middle and outer
	penumbrae of sunspots and that they stream out following paths traced by horizontal
	penumbral filaments (e.g., Bonet et al. \cite{bonet2004}, Sainz Dalda \&  Mart\'{i}nez Pillet \cite{sainz-pillet}, Ravindra \cite{ravindra2006}).  %\cite[e.g.][]{bonet2004, sainz-pillet,ravindra2006}.
	This has led to the hypothesis that they might be generated by
	Evershed flows (e.g., Cabrera Solana et al. \cite{cabsol2006}). More recently, by analyzing
	spectropolarimetric data, Kubo et al. (\cite{kubo2007}, \cite{kubo2007_2}) found
	that MMFs located along lines extrapolated from the horizontal component
	of the sunspot magnetic field display different properties from MMFs
	located along lines extrapolated from the vertical component. 

	Some MMFs, especially monopolar ones, are associated with bright
	features observed at various layers of the solar atmosphere
	(Harvey \& Harvey  \cite{harvey1973}, Hagenaar \& Frank \cite{hagenaar2006}, Lin et al. \cite{lin2006}, Choudhary \& Balasubramaniam \cite{choudhary2007}).
		Eruptive events, such as Ellerman Bombs (Nindos \& Zirin \cite{nindos1998}, Socas Navarro et al. \cite{socasnavarro2006}),
	microflares (Kano et al. \cite{kano2010}) or even CMEs (Zhang \& Wang \cite{zhang2002}), have also been reported
	in correspondance with MMFs.
	In addition, these features have recently been associated with shock waves traveling outward
	in the inner atmosphere (Lin et al. \cite{lin2006}, Socas Navarro et al. \cite{socasnavarro2006}, Ruytova \& Hagenaar \cite{ryutova2007}).

	Dipolar features have been interpreted as manifestations of $\Omega$ ( e.g., Schlichenmaier \cite{schlichenmaier2002}, Weiss et al. \cite{weiss04}) or U loops (e.g., Zhang et al. \cite{zhang2003}, Sainz Dalda $\&$ Bellot Rubio \cite{sainzdalda2008}, Kitiashvili et al. \cite{kitiashvili2010}) that travel along penumbral filaments.  Weiss et al. (\cite{weiss04}) have proposed that monopolar features of the same polarity of the parent spot are
flux tubes severed from the magnetic field of the spot. Those authors also interpreted monopolar features of opposite polarity with respect to the parent spot  as flux tubes
	that emerge in the photosphere at shallow angles.
	In addition, MMFs have been proposed to be
	manifestations of shocks (monopolar features) and solitons (dipolar features)
	traveling along the penumbral filaments (Ryutova \& Hagenaar \cite{ryutova2007}) .
	
	All these interpretations assume the presence of a penumbra or inclined magnetic filaments
	around the spot generating the MMFs. Besides, some authors have explicitly pointed out the need for a visible penumbra to generate these features (Mart\'{i}nez Pillet \cite{martinezpillet2002}, Zhang et a. \cite{zhang2003}).
	Nevertheless, MMFs have also been observed to stream out from spots without a penumbra. In particular,
	Harvey \& Harvey (\cite{harvey1973}) evinced from a statistical analysis of magnetograms that MMFs
	are associated with the presence of  moat regions rather than with penumbrae. 
	Zuccarello et al. (\cite{zuccarello2009}) observed MMFs streaming out from the naked spot
	\footnote{A spot that had developed and then lost a visible penumbra during its evolution.}
	of an active region during its decay phase. 
    Consistent with Harvey \& Harvey (\cite{harvey1973}),
	these authors also reported evidence of moat flows, whose connection with  penumbral filaments  had been debated in the past (e.g., Vargas et al. \cite{vargas2008}, or Deng et al. \cite{deng2007}) and whose independence from the presence of inclined
components of the magnetic field had been recently demonstrated by magnetic hydro 
dynamic (MHD) simulations (Rempel \cite{rempel2011}). This independence was also recently confirmed by Verma et al. (\cite{verma2012}), who observed MMFs of various types streaming from a naked spot and from a pore surrounded by moat flows. 

	In this paper, we report on the first high cadence (52~s), high spatial resolution ($\simeq~0.4$~arcsec), full-Stokes
observations of monopolar MMFs observed around a pore with a light bridge (for a description of the photospheric structure of a typical pore with a light bridge, see Giordano et al. \cite {giordano2008}) surrounded by a moat flow. This feature had developed a visible penumbra during days preceding our observations and should be therefore classified as naked spot. Nevertheless, for consistency with previous studies that have investigated the physical properties of this same feature (Stangalini et al. \cite{Stangalini2012}, Sobotka et al. \cite{sobotka2012}), we will address it as a pore  in the following. This choice is also motivated by the fact that the magnetic field of the structure does not exceed 2000 G (see Sec. \S 4), so that it can be classified as a pore according to the classification of Zwaan (\cite{Zwaan1987}).
	The paper is organized as follows: in \S~\ref{obs} and \S~\ref{ana} we describe the
	observations and computation of the relevant physical quantities;  in Section~\ref{evo} we describe the evolution
 of the observed active region and the detection of the investigated MMFs; 
	in Section~\ref{res} we present our results; in Section~\ref{disconc}
	we discuss and compare our findings with the recent literature;
	 conclusions are drawn in Section~\ref{sec:Conclusions}.  

\section{Observations and data reduction}
\label{obs}
	NOAA 11005 was observed close to disk center [25.2N, 10.0W] during its decay phase on October
	15th 2008 with IBIS (Interferometric BIdimensional Spectrometer, Cavallini \cite{cavallini2006}, Reardon \& Cavallini \cite{reardon2008}) at
	NSO/DST (National Solar Observatory, Dunn Solar Telescope).
	The dataset consists of $80$ scans of
	the \ion{Fe}{i}~$617.3$~nm and \ion{Ca}{ii}~$854.2$~nm lines acquired from 16:34:41~UT to 17:43:09~UT.
	Both lines were sampled at $21$~equally spaced spectral points.
	At each wavelength of the \ion{Fe}{i} line, the following
	six modulation states were acquired: $I+Q$, $I+V$, $I-Q$,
	$I-V$, $I-U$, and $I+U$. The field of view (FOV) is 40~$\times$~80~arcsec$^2$ with a pixel scale of
	$0.167$~arcsec~pixel$^{-1}$. White Light (WL, $621$~nm) and G-band
	($430$~nm) counterparts were acquired by imaging approximately 
	the same FOV of the spectropolarimetric data. 
	The pixel scales of the WL images and G-band images
	are $0.083$~arcsec~pixel$^{-1}$ and $0.051$~arcsec~pixel$^{-1}$, respectively.
	The exposure times for both the WL and spectropolarimetric images were 80 ms.
	The time needed to complete a spectropolarimetric scan was 52 s, which also sets the temporal resolution.\\
We applied the standard reduction pipeline (Judge et al. \cite{judge2010}, Viticchi\'{e} et al. \cite{viticchie2010})
	on the spectropolarimetric data, correcting for the instrumental
	blue-shift, the instrument, and the telescope-induced polarizations.\\	
During the acquisition, the high-order adaptive optics system of the
	NSO/DST (Rimmele \cite{Rim04}) was tracking on the pore,
	achieving good performance and stability.
	Due to the anisoplanatism, the good spatial resolution
	of the images degrades towards the edges of the FOV.
	In this study, we make use of a $\simeq36\times36$~arcsec$^2$
	part of IBIS FOV, approximately centered on the tracking point of the AO system,
	which shows excellent image quality.
	The residual atmospheric aberrations not corrected by the AO
	were estimated on WL and G-band images through the multi frame blind deconvolution technique
	(Van Noort, Rouppe van der Voort \& Lofdahl \cite{vannoort2005}) and then compensated for
	on the spectropolarimetric images through de-stretching. 
	After this post-processing, the estimated average resolution over the FOV is better than 0.4 arcsec at any time for the dataset duration.
\\

In order to reconstruct the history of NOAA 11005 we also employed other IBIS data acquired during the same campaign as well as MDI data and HINODE broad-band observations acquired from October 11th to October 17th 2008.

\section{Data analysis}
\label{ana}
	The magnetic flux along the line of sight (LOS) in each pixel was calculated by applying the center of gravity method (COG; Rees \& Semel \cite{rees1979})
	to the left- and right- circular polarization profiles acquired at the \ion{Fe}{i}~$617.3$~nm spectral range. Uitenbroek (\cite{uitenbroek2003}) proved that this method provides good estimates of the magnetic flux for field strengths up to $\approx$ 1 kG at the formation height of some \ion{Fe}{i} lines.

	Full-Stokes inversions of the points corresponding to the maximum value of the magnetic flux 
	associated with each MMF were carried out using the SIR code (Ruiz Cobo \& del Toro Iniesta \cite{ruiz1992}, Bellot Rubio \cite{bellotrubio2003}).
	    Given the average resolution of the spectropolarimetric dataset, the Stokes profiles were averaged
	 in a $3\times3$~pixel$^2$ box before the inversion
	procedure. The estimated polarimetric sensitivity  is  $2\times10^{-3}$ the continuum intensity level.	
	
	The inversions were performed by considering the polarimetric signals to be produced
	by a fraction $\alpha$ of magnetized plasma and by a fraction ($1-\alpha$) of field-free plasma. 
Magnetic field properties (i.e., strength, LOS, inclination, and azimuth in the plane perpendicular to the LOS) as well as the velocity of the plasma were assumed to be independent from height, whereas thermodynamical quantities were assumed height dependent. Both atmospheres were assumed to be described by a Milne Eddington approximation. An HSRA (Harvard-Smithsonian-Reference-Atmosphere, Gingerich et al. \cite{Gingerich71}) quiet Sun stratification
	was adopted for the initial-guess thermodynamical stratification of both components. 
		These inversion hypotheses do not allow us to interpret asymmetries in Stokes profiles and were
	adopted since our observations present a low degree of asymmetry. 
	The atomic parameters for the forward synthesis of the \ion{Fe}{i}~$617.3$~nm line were
	taken from  Norton et al. (\cite{norton2006}) and Barklem et al. (\cite{barklem2000}). Chemical abundances for iron were taken
	from Grevesse (\cite{grevesse1984}).
The typical uncertainties of values derived with SIR over the data inverted are 100 G, 0.3~km~s$^{-1}$, and 10 degrees for the magnetic field strength, LOS velocity, and LOS inclination, respectively. 

Inversions are known to suffer from uncertainties resulting from the ambiguity between the determination of field strength and the magnetic filling factor (e.g., Socas Navarro et al. \cite{socasnavarro2008}). Therefore, in order to validate our results, we also estimated the LOS magnetic field inclination and plasma LOS velocities by analyzing Stokes profiles. In particular, we computed the magnetic field inclination with respect to the LOS from the strong-field approximation
	($\theta_{SF}$, Landi degli Innocenti \& Landolfi \cite{landi2004}):
\begin{equation}
	\theta_{SF}=\arccos \frac{\sqrt{1+x^2} -1}{x}
	\label{form_inclin}
\end{equation}
	where $x = V/\sqrt{U^2 + Q^2}$ is computed at the left wing of the line. Beck et al. (\cite{beck2007}) showed this method to provide reasonable estimates of the magnetic field inclination in the photosphere, even when, as with our observations, the strong-field approximation (line Zeeman broadening much larger than the line thermal broadening) is not strictly verified.   
For those pixels whose Stokes $V$ lobes were higher than $3\times$ the polarimetric sensitivity, we also computed the LOS velocity from the shift of the zero-crossing of the \ion{Fe}{i} Stokes $V$ profiles.
LOS velocities in the chromosphere were estimated from the core shift of the \ion{Ca}{ii} line.
The zero-velocity references were estimated by the core positions of the time-and-space averaged line profiles in the quiet Sun part of the FOV. Inclination as well as photospheric and chromospheric LOS velocities were averaged on the same $3\times3$ pixel$^2$ box used for the inversions. 

\begin{figure*}
\centering

{\includegraphics[width=5.6cm,height=5.6cm, trim=1mm
1mm 2mm 8mm, clip]{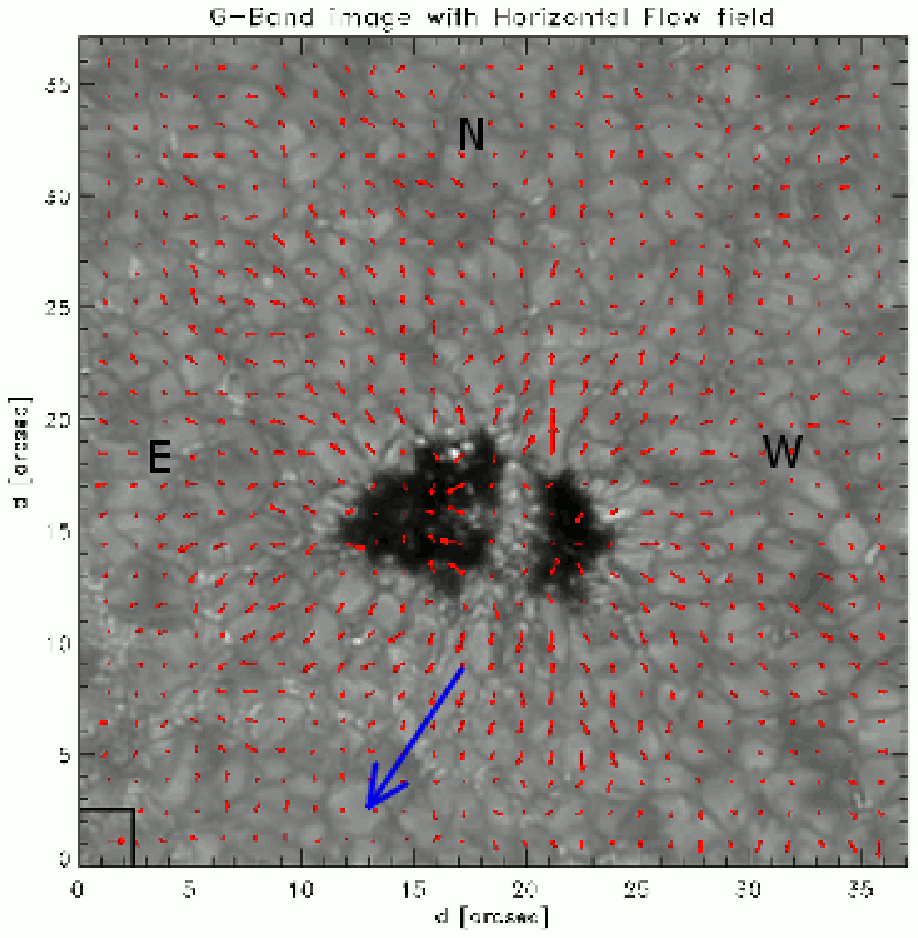}}
{\includegraphics[width=5.6cm, height=5.6cm, trim=1mm
1mm 5mm 20mm,clip]{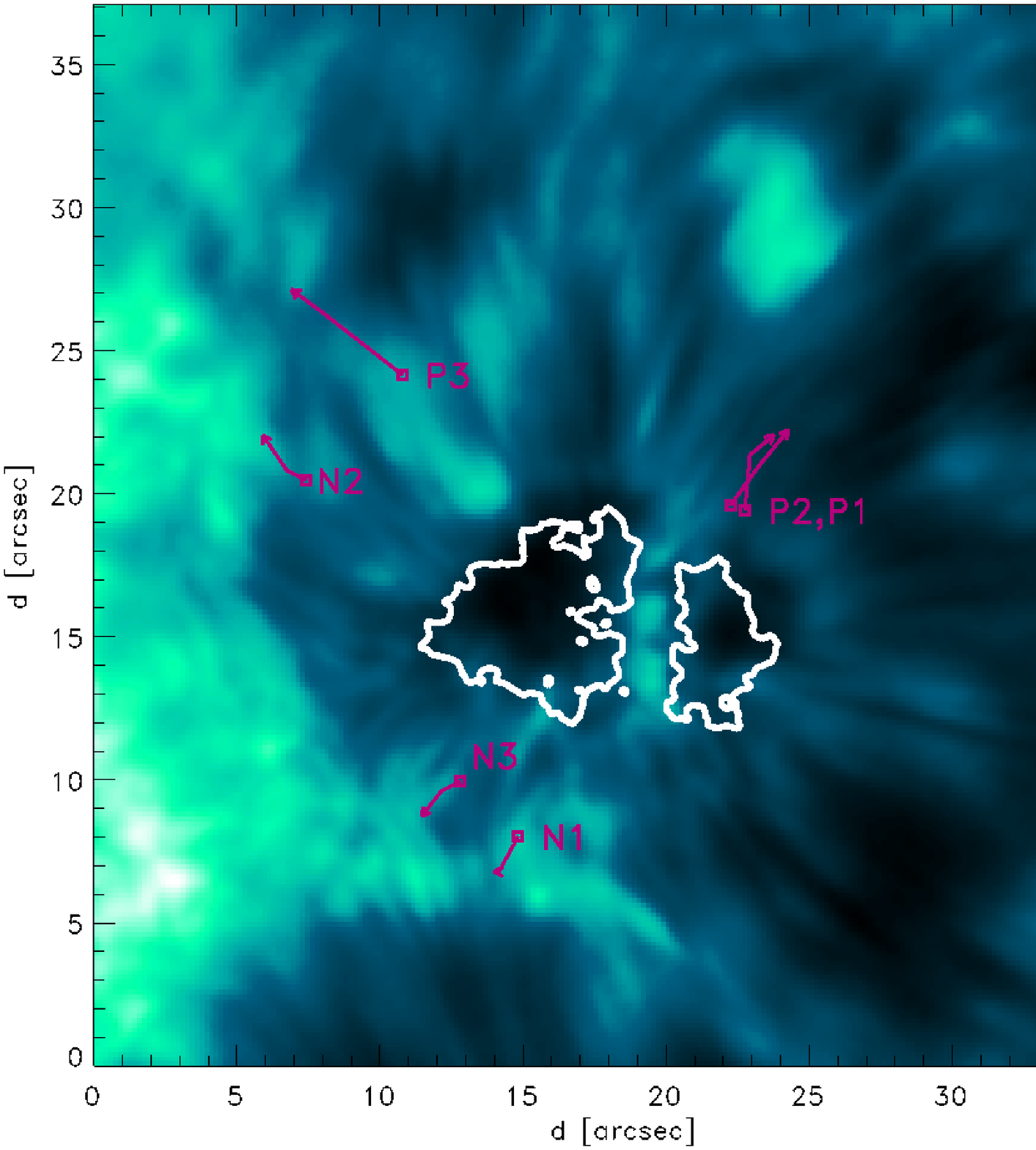}}
{\includegraphics[width=6.0cm, height=6.0cm,trim=0mm
1mm 3mm 1mm, clip]{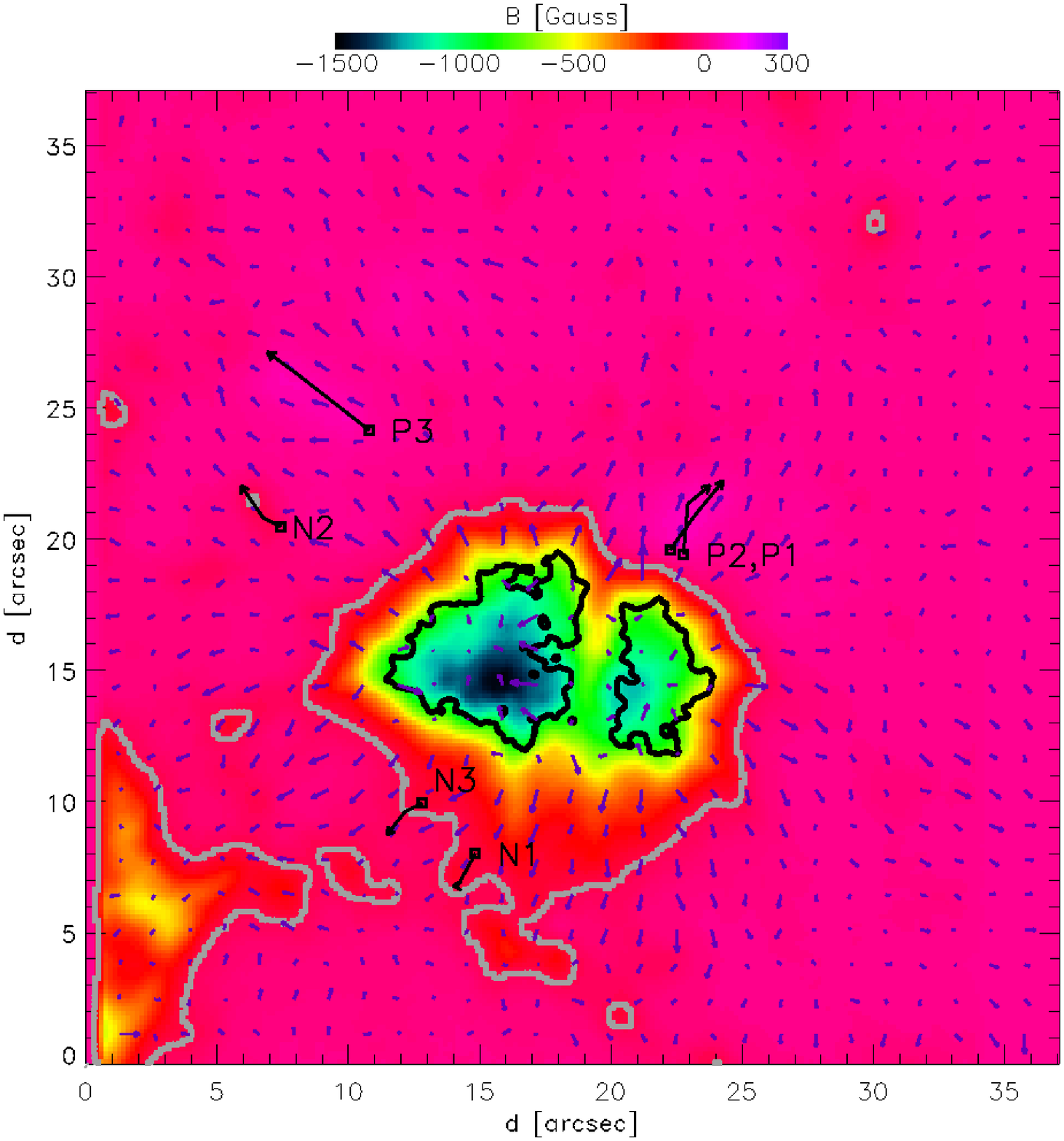}}

{\includegraphics[width=6.0cm,height=6.0cm,trim=0mm
1mm 3mm 1mm, clip]{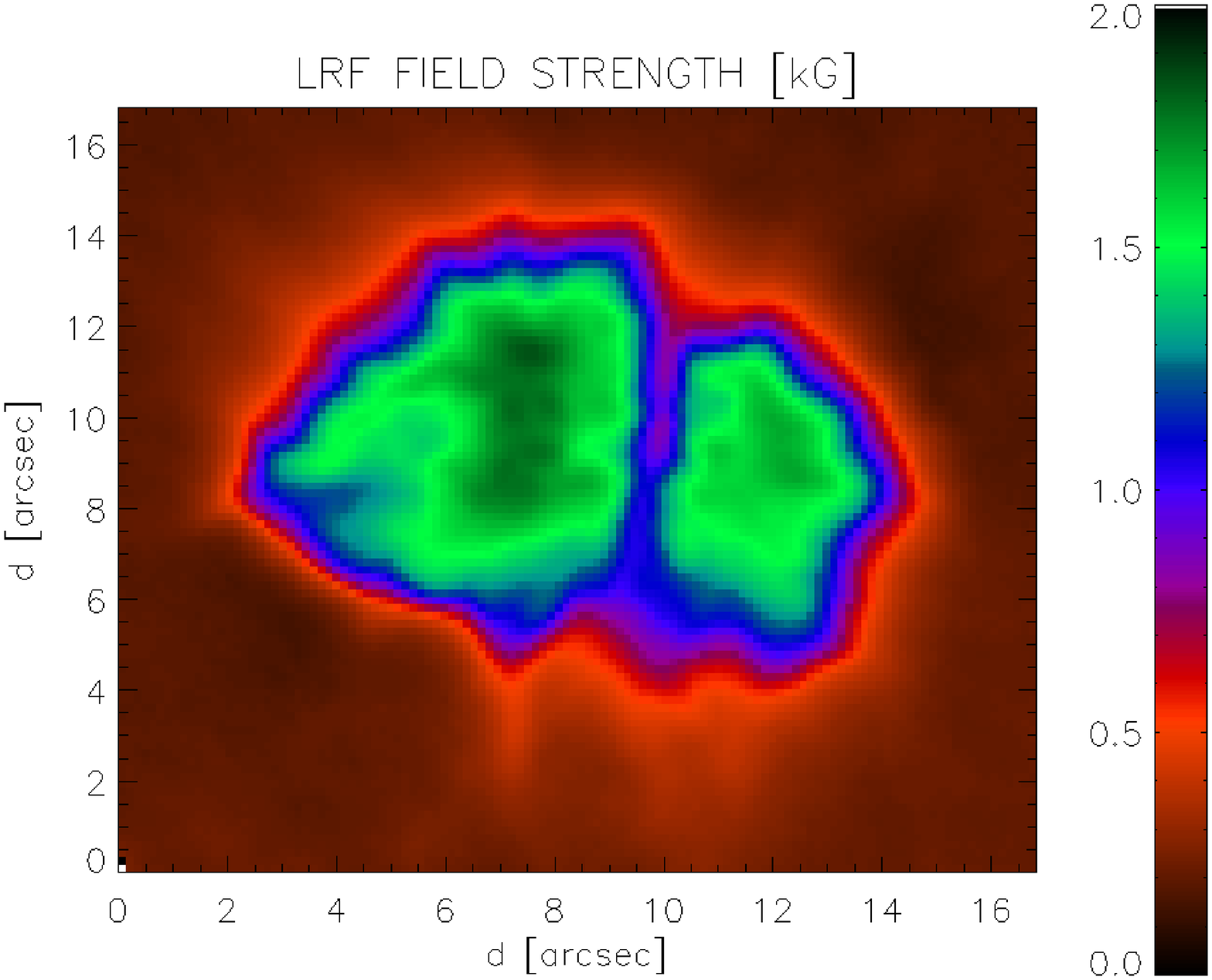}}
{\includegraphics[width=6.0cm,height=6.0cm,trim=0mm
1mm 3mm 1mm, clip]{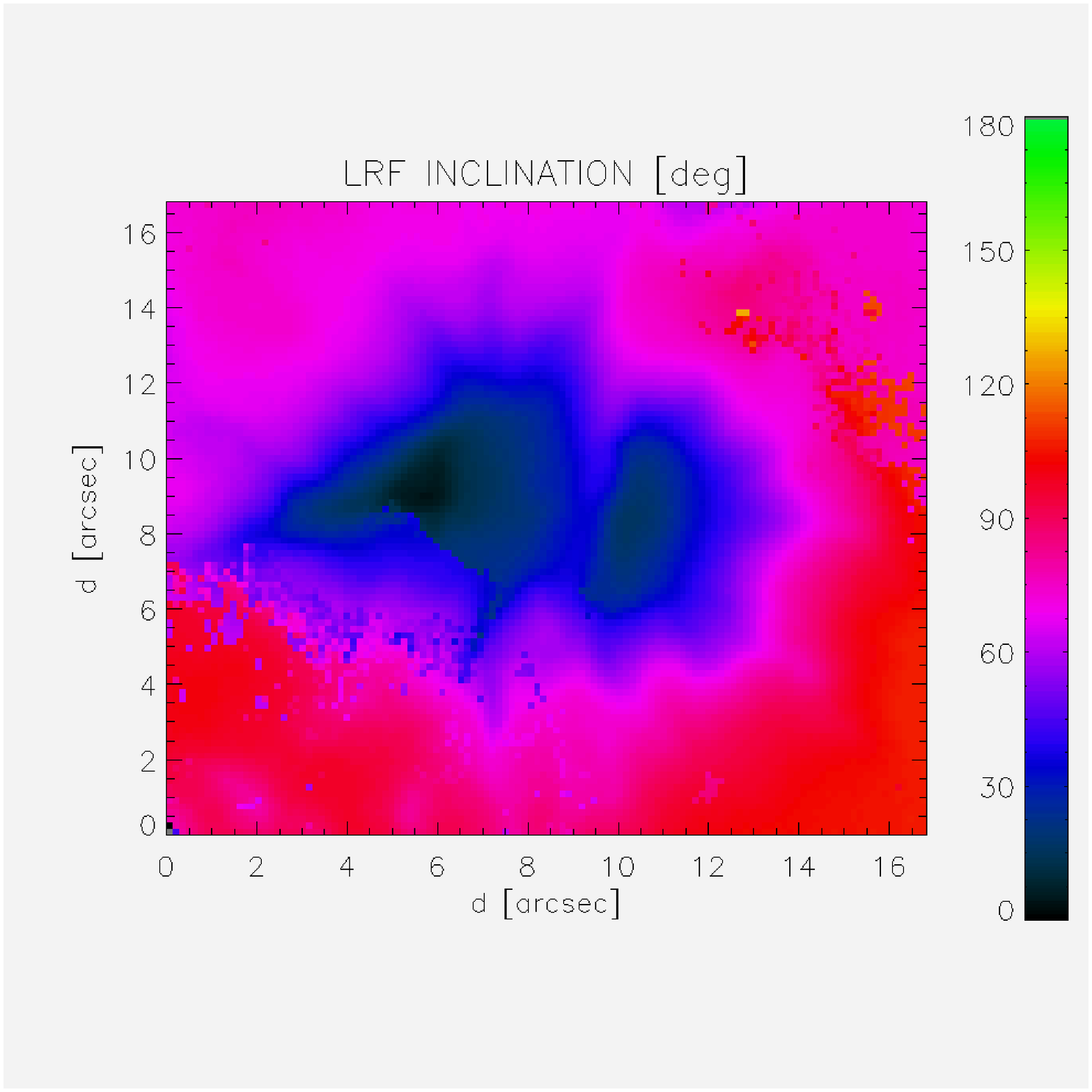}}
{\includegraphics[width=6.0cm,height=6.0cm,trim=0mm
1mm 3mm 1mm, clip]{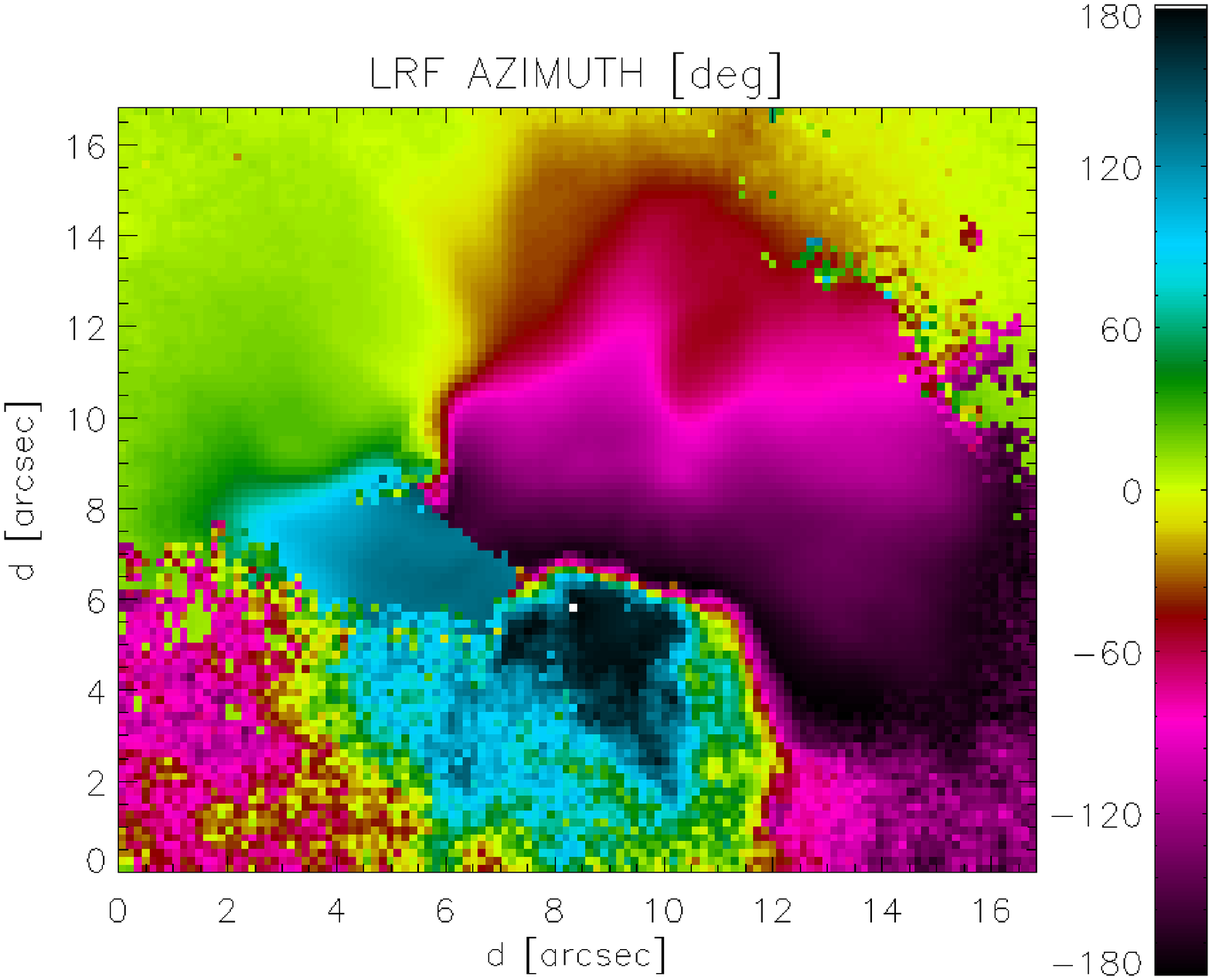}}

\caption{
Top left: LCT horizontal velocity field averaged over $70$~minutes superposed on a G-band image. The length of the red arrow in the lower-left corner of the image corresponds to a velocity of $1$~km~s$^{-1}$, while the black box shows the dimension of the correlation window used for the LCT computation. The blue arrow indicates approximate disk center direction.
Top middle: Corresponding FOV in the core of \ion{Ca}{ii} line.
Top right: Horizontal velocity field superposed on the median value (computed over $70$~minutes) of LOS magnetic flux; gray lines mark the regions where the magnetic flux is equal to -100 G. The little red (middle) and black (left) squares and arrows show the starting positions and trajectories of the six tracked MMFs. Bottom Left: Magnetic field strength. Bottom middle: inclination angle in the LRF. Bottom right:  azimuth angle in the LRF.
}
\label{fig_1}
\end{figure*}

	Horizontal velocities of the plasma were retrieved applying
	a local correlation tracking algorithm (LCT; November \& Simon \cite{november}) on G-band images.
	The correlation was computed on a $2.1\times2.1$~arcsec$^2$
	window (the small box shown in Fig. \ref{fig_1}, top left).

Finally, in order to evaluate the effects of the LOS direction on the observed geometry of the field, we analyzed SIR inversion results projected into the local reference frame (LRF) (Fig. \ref{fig_1}). These data derive from the study carried out by Sobotka et al. (2012) on the same region analyzed in this paper. We found that the differences of the inclination angle of the MMFs obtained in the two reference frames are comparable with uncertainties in SIR estimates (i.e., 10-20 degrees). Similarly, plasma velocities in the LRF were estimated considering the heliocentric angle of each pixel. Even in this case the velocity values obtained in the two reference frames differ within the uncertainty in the SIR estimates. As a result, we present in the following MMF properties obtained in the LOS reference frame only.

%------------------------------------------------------------------------------------------------------------------------------------------------
%--------------------------------------------------------------------------------------------------------------------------------------------------------
	
\section{Active region evolution and detection of MMFs}
\label{evo}
	 NOAA 11005 appeared on the solar disk on October 11th 2008 (as evinced by full-disk MDI magnetograms). In the first two days, it was composed of two
	sunspots of different sizes, separated by a light bridge (leading polarity),
	and a surrounding plage flecked with  small pores (following polarity).
	In subsequent days, the leading spot showed a partial penumbra (from MDI
	high-resolution magnetograms, HINODE continua, and IBIS October 13th observations).
	That penumbra disappeared before October 15th, leaving a large pore,
	suggesting that we observed NOAA 11005 during its decay phase. 
	In the meanwhile, the smallest of the pores grew in size and slowly
	merged with the largest one, while the following  small pores disappeared, 
	leaving only the plage region. On October 17th, NOAA 11005 had dissolved
	into two opposite polarity plages.

	On October 15th, IBIS FOV included the leading pore of the AR and
	part of the plage region of the same polarity as the pore, as illustrated in the detail
	of the full-disk MDI magnetogram shown in Fig.\ref{fig_appendix}.

	Examples of the analyzed IBIS FOV are illustrated in Fig. \ref{fig_1}. The top row shows a G-band frame (left), a \ion{Ca}{ii} core frame (middle), and  
	the median values over the temporal sequence of LOS magnetic flux (right). The bottom row shows the magnetic field intensity (left), inclination (middle), and azimuth (right) in the LRF, derived from SIR inversions, as described in Sobotka et al. \cite{sobotka2012}.
	These figures show that the pore is surrounded by an area of magnetized plasma, whose flux decreases and whose inclination increases with the distance from the umbra. Similar magnetic configurations around pores have been reported by other authors and are usually referred to as magnetic canopy of the pore (e.g., Keppens \& Mart\'{i}nez Pillet \cite{keppenspillet1996}, Leka \& Steiner \cite{lekasteiner2001}, Bellot Rubio et al. \cite{bellotrubio2008}).  Moreover, both photospheric inclination map and \ion{Ca}{ii} frame show that the lines of the field are inclined on the side farthest
	from the same-polarity plage region (see Stangalini et al. \cite{Stangalini2012} and Sobotka et al. \cite{sobotka2012} for a detailed description of the field geometry of this magnetic region). 
	Probably, this signature is the remnant of the partial penumbra observed in
	the days previous to our observations.  

Figure \ref{fig_1} also shows the LCT horizontal velocity field averaged over the whole dataset duration ($\simeq70$~minutes). 
A global outward motion from the pore (moat flow) that extends up $\sim 10$~Mm from the umbra boundary is clearly observed (see also Sobotka et al. \cite{sobotka2012}).
	All the MMFs we observed appeared in this region, but their motions can not be totally ascribed to the outward-flows
	of the photospheric plasma, as described in the following paragraphs.\\

 Figure \ref{fig_1} also shows the trajectories of the six
MMFs investigated in this study.  Their evolution can be observed  in the movie  reported in the online material, which shows the LOS magnetic flux  clipped at -200 G and 100 G. The movie shows the presence of
	several small features of both polarities escaping the pore; among these we investigated the most evident and longest living. The analyzed features were labelled P or N, depending whether their polarity was opposite or the same with respect to the polarity of the pore. It is important to notice that the detection of MMFs is hampered by the presence of the magnetic field of the parent feature (spot or pore) that they stream from (Kubo et al. \cite{kubo2007}). This is because they can be detected only when MMFs' flux exceeds that of the parent feature and/or when they are far enough not to be embedded anymore in the parent magnetic field. For the same reason, the definition of the time and location of an MMF formation can be tricky (Zhang et al. \cite{zhang2003}). In our case, we have identified the N features on magnetograms clipped at -200 G. 
The choice of this threshold sets both the moment of detachment and location for N1 and N3, which are closest to the pore. P1 and P2, the closest to the pore among P features, have opposite polarity and are thus not prone to this kind of uncertainty, although we cannot rule out the possibility that they formed before and that we could detect them only when their flux exceeded that of the pore. Nevertheless, the images in Fig. \ref{fig_1} clearly show that P1, P2, N1, and N3 form at the borders of the magnetic field of the pore, following outward trajectories.  
P3 and N2, however, formed before we started our observations and are the farthest from the pore that we observed. On the other hand, their trajectories suggest that they also originated from the pore field.

\section{Evolution of the MMFs}
\label{res}
   	Six MMFs have been tracked to extract their trajectories and velocities.
    To track them, we used the position of their associated maximum values on the LOS magnetogram.
    Whenever a possible ambiguity arose, we used the integrated polarization signal to retrieve the same information.
	We investigated the evolution of several physical quantities
	of these MMFs, either directly computed from the
	spectropolarimetric data or obtained from the SIR inversion. 
\subsection{Type P features}
\label{resIII}
%+++++++++++++++++++++++++++++++++++++++++++++++++++++++++++++++++++++++++++++
% Figure 1 P1
\begin{figure*}[!t]
\centering
\includegraphics[width=17cm,trim=13mm 0mm 6mm 0mm]{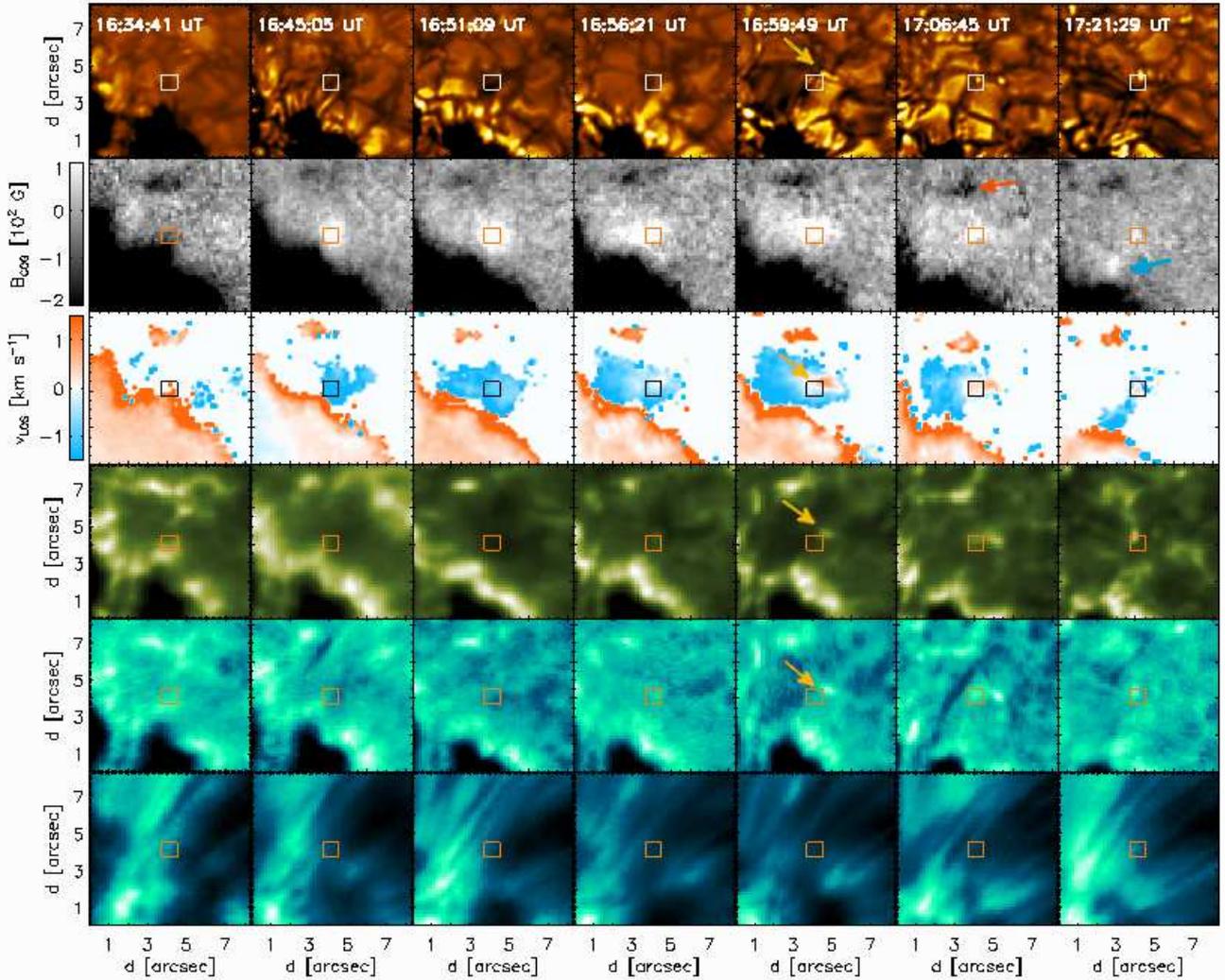}
  	\caption{Representative $8\times8$~arcsec$^2$ frames extracted from the complete temporal
  	evolution of P1.
  	 From the top to
	the bottom: G-band filtergrams; LOS magnetograms (clipped at $+100$ and $-200$~G) derived with COG;
  	LOS velocity fields as derived from the Stokes~$V$ zero-crossing point
  	(clipped at $\pm1.5$~km~s$^{-1}$); \ion{Fe}{i} core images; \ion{Ca}{ii}
  	wing images; \ion{Ca}{II} core images. The red box is centered on the maximum of the LOS magnetic flux within P1. Yellow arrows in the fifth column mark the formation of bright features in filtergrams and G-band frames and the corresponding formation of a downflow within the MMF. The red arrow on the magnetogram of the sixth column marks a small magnetic region of opposite polarity with respect to P1. The blue arrow on the LOS magnetogram of the last column marks P2. Frames from the second column on
  	correspond to the times marked by arrows in Fig.~\ref{fig_p1evo}.
  	\label{fig_p1text}}
\end{figure*}

Figure \ref{fig_1} shows that	P features originate at the north of the pore, where the field is highly inclined.	
They migrate preferentially in the opposite direction with respect
	to the plage region, following  trajectories  that resemble those 
 of overlying chromospheric fibrils. The average horizontal speed for the three features is $\sim$ 1~km~s$^{-1}$, although they decelerate during their evolution. They appear as loose features of sizes of a few arcsec$^{2}$, whose coherence varies with time. 

We could observe the formation, evolution, and disappearance of P1 and P2 .
These phases are illustrated  in Fig.~\ref{fig_p1text} for P1. The red box is centered on the maximum of the magnetic flux within the MMF. 
	The first column shows
	the environment at the location of P1 (the square box is located where the MMF will appear) just a few minutes before
	its formation. The magnetic feature originates in a
	region of strong magnetic and velocity gradients at the outer boundary of the canopy.

	The second column shows the phase in
	which P1 is clearly separate from the field of the pore. We note that a region of upflowing magnetized plasma  corresponds to the location of the MMF. 
	As the MMF evolves, it first increases in size
	(third column) and then loses spatial coherence (fourth column on). This loss of coherence coincides with
	the formation, approximately 15 minutes after the appearance of P1, of downflows within P1 and with the appearance of positive contrast features in photospheric observations (yellow arrows, fifth column). We notice that these positive contrast features firstly appear as elongated structures within an intergranular lane and later on evolve into two easily distinguishable bright points,
	 corresponding to the highest magnetic flux signals within the MMF (sixth column). 
	 No clear intensity  enhancement is observed in \ion{Ca}{ii} core images. 
 
	The last column shows the phase in which the magnetic flux associated with P1
	is very close to the noise level, so that one can barely identify the MMF.
It also shows the formation of P2 (blue arrow), which occurred almost at the same location as P1 and followed
	a very similar trajectory. Similar to P1, P2 had associated upflow plasma motions at its appearance, and brightening in the photosphere occurred approximately 15 minutes after its formation.

All the type P features investigated migrate toward  small magnetic patches of opposite polarity (red arrow in the sixth column in the case of P1) with which they never merge. In the case of P3, the Ca II wing and core images revealed  enhancements of contrast in the region between the MMF and the opposite polarity patch that might indicate reconnection events, as described by Kano et al. (\cite{kano2010}). 
		
\begin{figure}%[!t]
\centering
\includegraphics[width=16cm,trim=3mm 15mm 7mm 5mm   ]{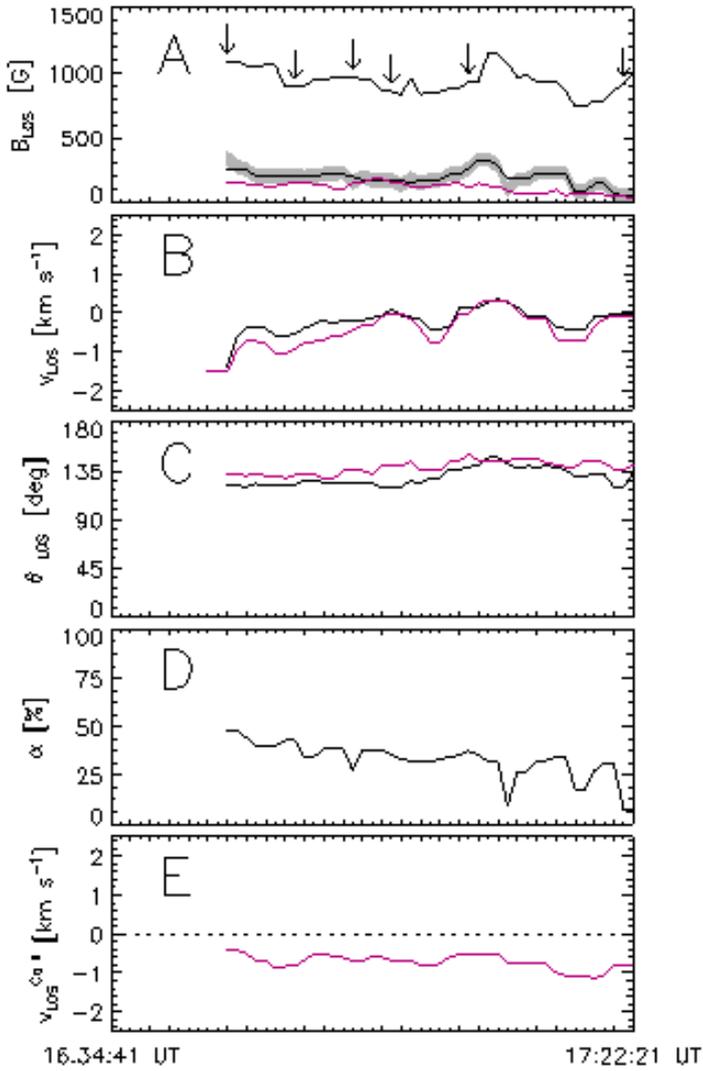}
  	\caption{Temporal evolution of the physical properties of P1.
  	\textit{Panel A}: magnetic field intensity derived from the
  	inversions (black line); LOS magnetic flux density from COG (magenta line) and from the inversions (black thin line); the shadow represents
  	a range of $\pm50$~G around the magnetic flux density values from the inversions. \textit{Panel B}: photospheric LOS velocity derived
  	from the inversions (black line) and from the Stokes $V$
  	zero-crossing (magenta line). \textit{Panel C}: LOS magnetic field inclination derived from the inversions (black line)
  	and from the strong field approximation
  	(magenta line). \textit{Panel D}: magnetic filling factor derived
  	from the inversions. \textit{Panel E}: chromospheric LOS velocity derived from
  	the core shift of \ion{Ca}{ii} line. 
	The arrows mark the times corresponding to the frames shown in Fig.~\ref{fig_p1text} from second column on. Minor horizontal tick marks are spaced 52 s.
  	\label{fig_p1evo}}
\end{figure}

	% Inversion
	Figure~\ref{fig_p1evo} shows the evolution of physical properties of P1 derived from inversions and analyses of Stokes profiles. First of all, we notice that there is a general agreement between results obtained with the various techniques. In particular, the LOS magnetic flux density derived from inversions 
is consistent with the value
	derived with COG, as it varies around $200$~G in both cases. 
	The average magnetic field intensity along the sequence is about $1$~kG, and it slowly decreases with the increase of the distance of P1 from the pore.
	The LOS velocity is upflow at the appearance of the
	MMF. Afterwards, a sequence of upflows and downflows is observed, with amplitudes that never exceed $1.5$~km~s$^{-1}$. As an analysis of the 
power spectra of the LOS velocities did not show power in the 2.8-3.8 mHz range, we exclude a significant contribution of the p-modes to the observed variations.	

	The LOS inclination of the magnetic field decreases almost monotonically with time, ranging approximately  from $120$ to $145$~degrees for values derived from inversions and from $130$ to $150$~degrees for values derived from strong field approximation. 
 
	The magnetic filling factor also decreases with time,  in agreement with the already described loss of spatial coherence of P1. 

	As for the chromospheric LOS velocities, the values derived from the \ion{Ca}{ii} core doppler shift do not
	reveal any clear relation with the photospheric LOS velocities. In fact, upflow motions of approximately $-1$~km~s$^{-1}$ are consistently measured along the observations.

The evolution of physical properties of P2 is very similar to that of P1. The evolution of P3 did not show any particular trend; average values of measured quantities for this feature are $1.2$~kG, $200$~G, $30$~degrees, $-1.5$~km~s$^{-1}$, and  $-1.$~km~s$^{-1}$ for field strength, magnetic flux, LOS inclination,
and LOS photospheric and chromospheric velocities, respectively.  

%+++++++++++++++++++++++++++++++++++++++++++++++++++++++++++++++++++++++++++++
% Figure 2 P1
%+++++++++++++++++++++++++++++++++++++++++++++++++++++++++++++++++++++++++++++

\subsection{Type N features}
\label{resII}

%+++++++++++++++++++++++++++++++++++++++++++++++++++++++++++++++++++++
Figure \ref{fig_1} shows that N1 and N3 form in a region near the pore, where the magnetic field is mostly directed along the LOS. However, the trajectory of N2
is compatible with a formation in a region where the field of the pore is more inclined.  
With respect to type P features, type N features are smaller, more compact, and travel at lower speeds (few hundreds of m~s$^{-1}$), following approximately radial trajectories. N1 and N3 are embedded in the field of the pore, and their dynamics are strongly influenced by the interaction with same polarity features  nearby.  The evolution of N2 is more regular, as it is the farthest from the pore and does not interact with other magnetic features. 
 Consistently, all the
tracked type N features slow down after traveling a certain distance, reaching regions
where magnetic fields of the same polarity are accumulated.   	

Both N1 and N3 detach from the canopy some minutes after the beginning of our observations.  Nevertheless, in correspondence of the regions where these MMFs will form, bright features are present in the G-band, \ion{Ca}{ii} wing, and \ion{Fe}{i} core frames from the beginning of the observations. After the appearance of the MMFs, these bright features travel following the same trajectory as the MMFs and, like type P, evolve with time in both shape and number.
	Inspection of photospheric dopplergrams showed that type N features have mostly associated downflow plasma motions. 
The \ion{Ca}{ii} core frames show very different characteristics from those
observed in correspondence of the type P features. The most evident one is the lack of arches and fibrils for N1 and N3, whereas fibril structures are observed above N2, in agreement with the fact that this feature evolves in a region where the field surrounding the pore has an inclined component. In contrast to type P features, intermittent brightening is observed in Ca II core frames. The analysis of COG magnetograms showed that both magnetic flux density and the size of N1 and N3 vary with time, with their
increases sometimes associated with the merging with other magnetic features. 

\begin{figure}[!t]
\centering
\includegraphics[width=15cm,trim=3mm 15mm 7mm 5mm ]{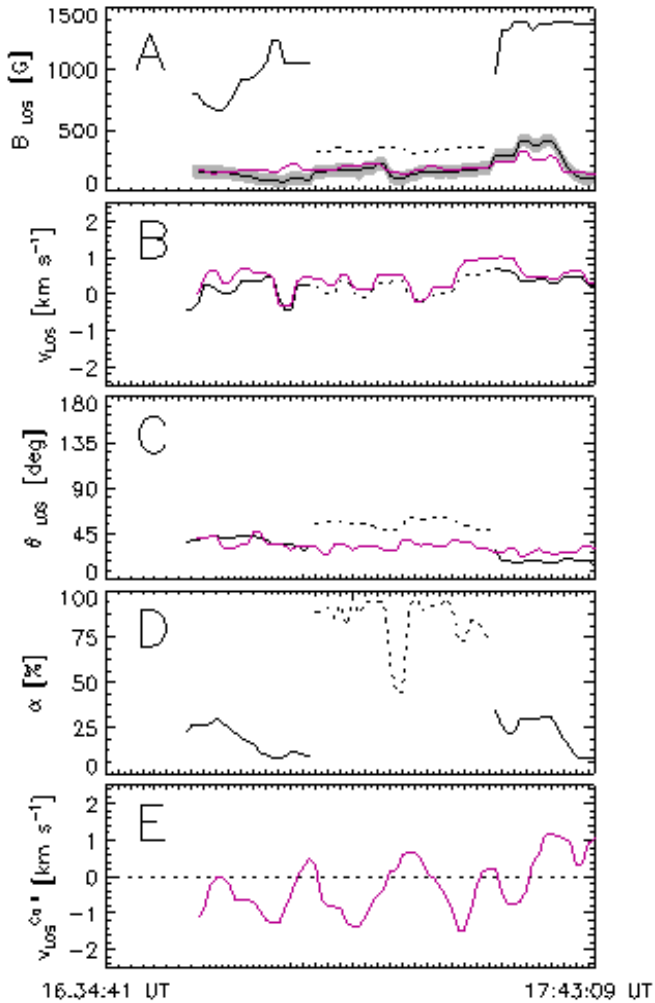}
  	\caption{Temporal evolution of the physical properties of N1.
  Legend	as in Fig.~\ref{fig_p1evo}. Values from inversions indicated with dashed lines are considered not reliable (see text).\label{fig_n1evo} }
\end{figure}

An example of quantitative analysis of properties of type N features is reported in Fig. \ref{fig_n1evo}, that shows the results obtained for N1. Unlike type P features, the temporal coherence of results obtained from inversions is not preserved along the sequence, as sharp variations of the derived values (indicated with dashed lines) occur. At those times, the field strength drops to few hundreds Gauss and discrepancies with values derived with other techniques are also observed. This suggests that those results are not reliable, although the magnetic flux values derived from inversions are still consistent with values derived from COG. 
As mentioned in Section~\ref{obs} this is because
the ambiguity problem between
the magnetic field strength and the magnetic filling factor cannot be avoided. 
 In the following, we thus reject results from inversions for which the magnetic field strength is below $500$~G.	 
	With the exception of these points, Fig. \ref{fig_n1evo} shows that the field strength
	varies with time, mostly because of merging with features of the same polarity. 
 The value of the inclination angle decreases with time, indicating that the field of N1 becomes
	more vertical as the feature moves away from the pore. 
	We also note that N1 has associated photospheric downflow plasma motions during the first minutes after its detachment and that later upflow and downflow motions
	alternate along the sequence.
    Even in this case, the LOS velocity power in the 2.8-3.8 mHz range is very low, thus excluding
    a relevant contribution from the p-modes.	
Chromospheric LOS velocity derived from the \ion{Ca}{ii} core shift also shows variations on various time scales from upflows to downflows. This fact, together with the variable intensity enhancement observed in \ion{Ca}{ii} frames, might suggest the presence of traveling waves (Lin et al. \cite{lin2006}). This issue will be investigated in a subsequent study.

The physical evolution of N3 is similar to that of N1, although it is much shorter-lived. Physical properties of N2 are also similar to those estimated for
N1, with the exception of the LOS photospheric velocity, that was consistently downflow and had an amplitude of $\approx$ $2$~km~s$^{-1}$ along the whole sequence. 

%+++++++++++++++++++++++++++++++++++++++++++++++++++++++++++++++++++++
% Figure 2 N1
%+++++++++++++++++++++++++++++++++++++++++++++++++++++++++++++++++++++

\section{Discussion}
\label{disconc}
The physical properties of MMFs derived from our analysis are summarized in Section \ref{sec:Conclusions}.  
They generally agree with results presented in the literature for MMFs streaming out from sunspots with penumbrae.	
In this section, we discuss and compare our results with those from spectropolarimetric observations recently presented in the literature.

 First of all, we notice that type P features appear (or their trajectories are consistent with an origin) at the borders of the pore canopy, in a region where the magnetic lines are highly inclined. Conversely, type N features stream (or their trajectories are consistent with an origin) from regions at the borders of the canopy
	where the field is almost vertical. A similar correlation
	between the polarity of MMFs and the magnetic field inclination of the parent spot was
	already reported by Cabrera Solana et al. (\cite{cabsol2006}) and Kubo et al. (\cite{kubo2007,kubo2007_2}) for features
	streaming from penumbrae.

\begin{figure}
\centering
\includegraphics[ trim=10mm 5mm 0mm 0mm]{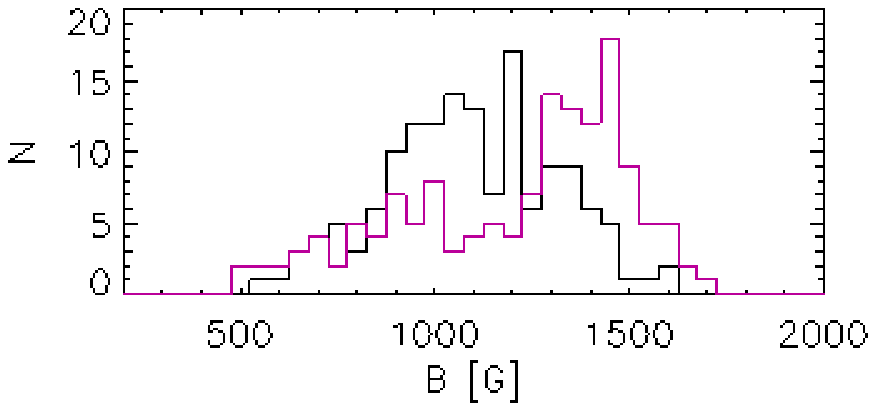}
\includegraphics[ trim=10mm 5mm 0mm 0mm]{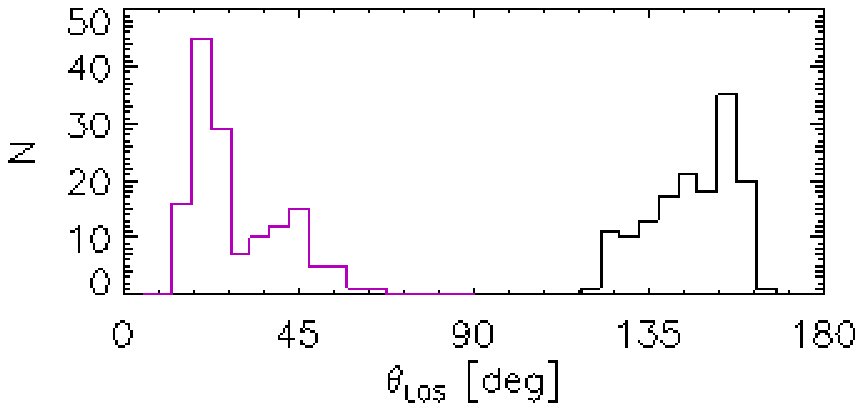}
\includegraphics[ trim=10mm 0mm 0mm 0mm]{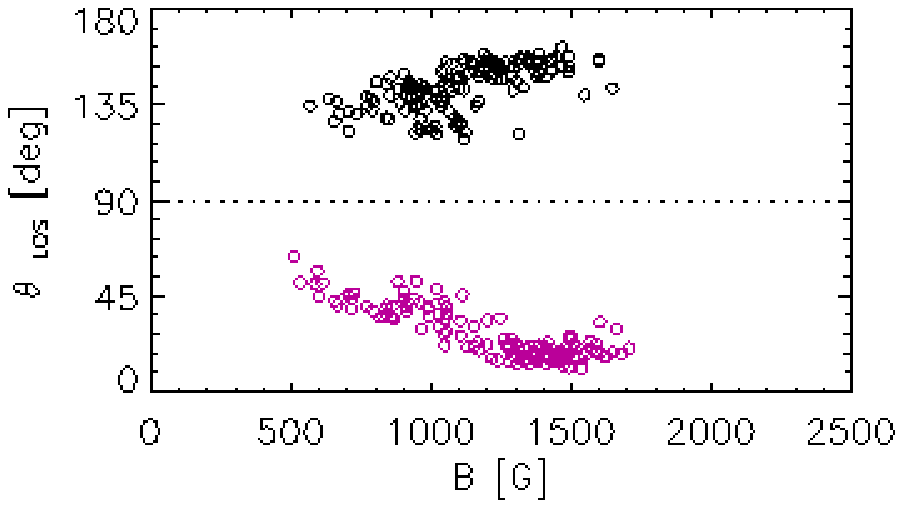}
  	\caption{\textit{Top}: distributions of absolute magnetic field strength values estimated with SIR inversions. \textit{Middle}:
distributions of LOS magnetic field inclination values estimated with SIR inversions.
\textit{Bottom}: scatter plots of absolute magnetic field strength vs inclination. In all panels, black symbols indicate type P features and magenta symbols indicate type N features. \label{stat}}
\end{figure}

	 Figure~\ref{stat} shows the distributions of the magnetic field
	intensity (top) and LOS inclination values (middle) obtained from inversions at different instants for the six features. Note that the shapes of these distributions are determined by the evolution of the MMFs and, most importantly,  that the number of investigated features is too exiguous to infer their statistical properties. Nevertheless, because different features can be observed at different phases of their evolution during an observation, the derived distributions provide indications about the most likely configurations during observations of features with properties similar to the ones we observed. 
We note that the distributions of the magnetic field intensity are widely spread around B $\simeq$ 1.2 kG for both types of MMFs, although the most probable value found for type N (1.5 kG) is higher than the most probable value retrieved for type P features (1.3 kG). These values are consistent with those reported by Kubo et al. (\cite{kubo2007}), although they did not report any significant correlation between the field strength and the polarity of the feature.  Instead, Choudary \& Balasubramaniam (\cite{choudhary2007}) found that a larger magnetic field is on average associated with MMFs of polarity opposite to that of the parent spot, although they reported that the typical field strengths of both types of MMFs were below kG values with an average filling factor between 30-40\%.

The distributions of the magnetic field inclination values show that for both type P and N features the magnetic field vector can be strongly inclined (i.e., up to $\simeq120$~degree for 
	type P and up to $\simeq50$~degree for type N), but the most probable configurations are almost vertical for both types.

The bottom panels in Fig.~\ref{stat} show the inclination angle of the field versus its (absolute) strength.
The plots show that the inclination varies almost linearly with the field strength and that the larger the field the more vertical the direction.
A similar relation had been already reported by Kubo et al. (\cite{kubo2007}).

Our results also show that the two types of features are characterized by different LOS photospheric plasma motions.
Upflow motions are found for type P features, especially during the first stages of their evolution. Downflow plasma motions are found for type N features. The velocity values we found are comparable with typical convective motions, since they do not exceed $2$~km~s~$^{-1}$. To the best of our knowledge, the literature lacks observations about LOS velocity of monopolar MMFs in the photosphere. As a result, we are not able to compare this result with previous observations.  

%+++++++++++++++++++++++++++++++++++++++++++++++++++++++++++++++++++++++++++++++

%+++++++++++++++++++++++++++++++++++++++++++++++++++++++++++++++++++++++++++++++

\section{Conclusions}
\label{sec:Conclusions}
We analyzed IBIS spectropolarimetric data of monopolar MMFs 
observed around a pore that had lost its penumbra.
From the evolution of six tracked MMFs, we found that they share the
following characteristics:

\begin{itemize}
\item 
They move in a moat flow.
\item 
Type P features are generated in a region where the magnetic field surrounding the pore is highly inclined, whereas type N features stream from a region where the field surrounding the pore is vertical.  
\item 
Type P are bigger than type N features, but less spatially coherent.
\item
Type P features move at average horizontal speeds of $\approx$ 1 km~s$^{-1}$,  while type N features are slower (hundreds of m~s$^{-1}$);
 both types decelerate with the increase of the distance from the pore.
\item
LOS plasma motions within the MMFs are mostly upflow for type P and downflow for type N features. 
\item 
For type P features, the appearance of bright structures in photospheric images occurs
approximately 15 minutes after the formation of the MMFs. 
For type N features, bright structures in photospheric images are observed before their detachment from the pore field.  
\item Variable brightenings in the \ion{Ca}{ii} core frames are observed in correspondence of type N features. 
\item
The absolute value of the magnetic flux associated with MMFs is within 100 and 250~G. 
\item 
The magnetic field strength distribution ranges from 500 G to 1700 G. The most likely values are 1.3 and 1.5 kG for type P and type N features, respectively.
\item
The inclination of the magnetic field varies with time, although the most likely configuration is almost vertical for both types of features, with smaller inclination found for type N features.
\end{itemize}

All these properties are commensurate with the properties of monopolar MMFs streaming from spots with penumbrae. We conclude that they are
very likely the manifestation of the same physical processes (as  also recently suggested by Sainz Dalda et al. \cite{sainz2012}).

Ultimately, the main peculiarity of our observations with respect to previous studies is the high temporal cadence ($52$~s) of the data employed, which 
	allowed us to investigate the temporal evolution of the physical properties of the
	MMFs observed. The evolution of the two type N features closest to the pore is driven by the surrounding environment, as these MMFs are always embedded in the outskirts of the parent pore, where new flux emerges or is expelled. For these features,  the appearance of
	the photospheric brightening precedes the formation of the MMF. The MMF farthest from the pore travels away in a more regular trajectory. These results
	suggest that type N features are magnetic flux bundles expelled from the boundary of
	the pore. During the evolution of type P
	MMFs, we found instead a slow decrease of both the magnetic flux and the upward LOS plasma motions, while the field becomes more vertical the larger the distance travelled from the pore. The photospheric brightening is delayed with
	respect to the appearance of the MMFs. Similar observational evidence has been recently
	interpreted as magnetic arches that rise from the photosphere to the higher layers of the
	atmosphere (Mart\'{i}nez Gonz\'{a}lez \& Bellot Rubio \cite{martinez2009}), thus suggesting that type P MMFs might be the manifestation
	of a similar physical process. 

 Similar interpretations on the nature of monopolar MMFs have been already suggested (e.g., Weiss et al. \cite{weiss04}, Kubo et al. \cite{kubo2007}).
Results from numerical simulations that can confirm these interpretations, as well as explain in general the physical nature of the various types of MMFs  and reproduce their observed physical properties have not yet been presented in the literature.

\begin{acknowledgements}
This study was supported by the Istituto Nazionale di
Astrofisica (PRIN-INAF-07/10), the Ministero degli Affari Esteri
(Bando 2007 Rapporti bilaterali Italia-USA), and the Agenzia
Spaziale Italiana (ASI/ESS/I/915/01510710). The NSO is operated
by the Association of Universities for Research in Astronomy,
Inc. (AURA), for the National Science Foundation. IBIS was
built by INAF/Osservatorio Astrofisico di Arcetri with
contributions from the universities of Firenze and Roma ``Tor Vergata,''
the National Solar Observatory, and the Italian Ministries
of Research (MIUR) and Foreign Affairs (MAE).  S.C., I.E., F.Z., and F.B. thank the
International Space Science Institute
for the opportunity to discuss this topic
at the international team
meetings "Filamentary Structure and Dynamics of Solar Magnetic Fields" and "Magnetic Flux Emergence in the Solar Atmosphere."
We are grateful to A. Tritschler and S. Guglielmino for the fruitful comments and suggestions. We are also grateful to M. Sobotka for providing data of the magnetic field in the local reference frame. 
\end{acknowledgements}

\Online

\begin{appendix}

\section{MDI magnetogram}
\begin{figure*}[h!]
\centering{
\includegraphics[width=11.5cm, height=11.5cm, trim=2mm
4mm 0mm 6mm,clip]{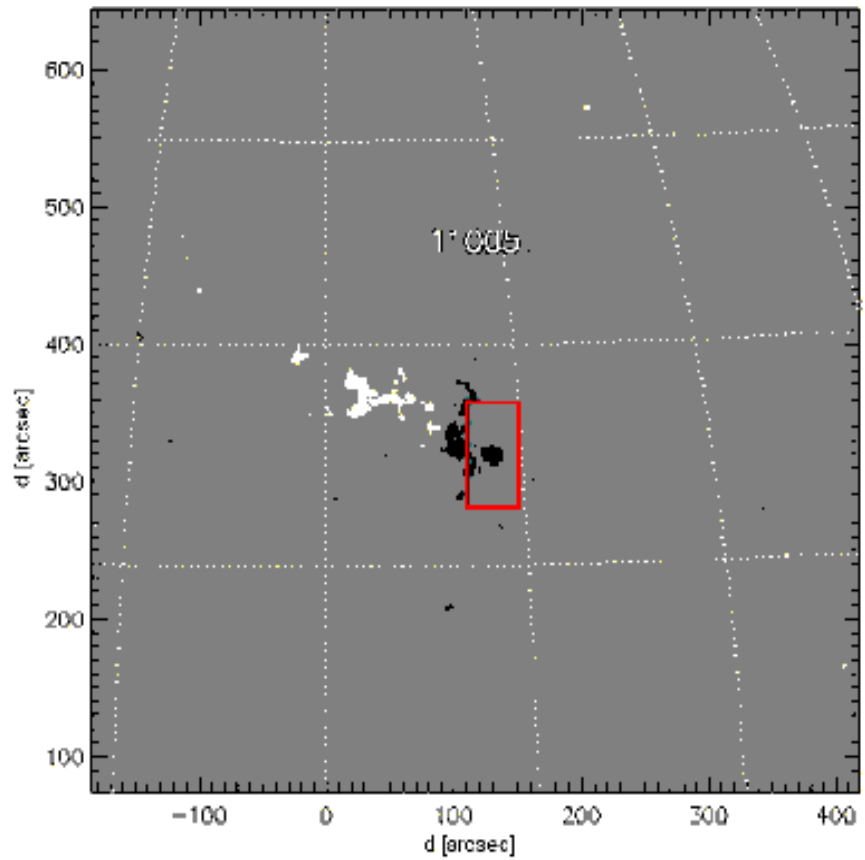}
}
\caption{Detail of a full disk MDI magnetogram acquired on October 15th 2008, 11:15 UT. The red box shows the approximate FOV of IBIS.
}
\label{fig_appendix}
\end{figure*}

\section{Online movie}
\begin{figure*}[!h]
\centering{
\includegraphics[width=11.cm, trim=19mm
5mm 15mm 5mm,clip]{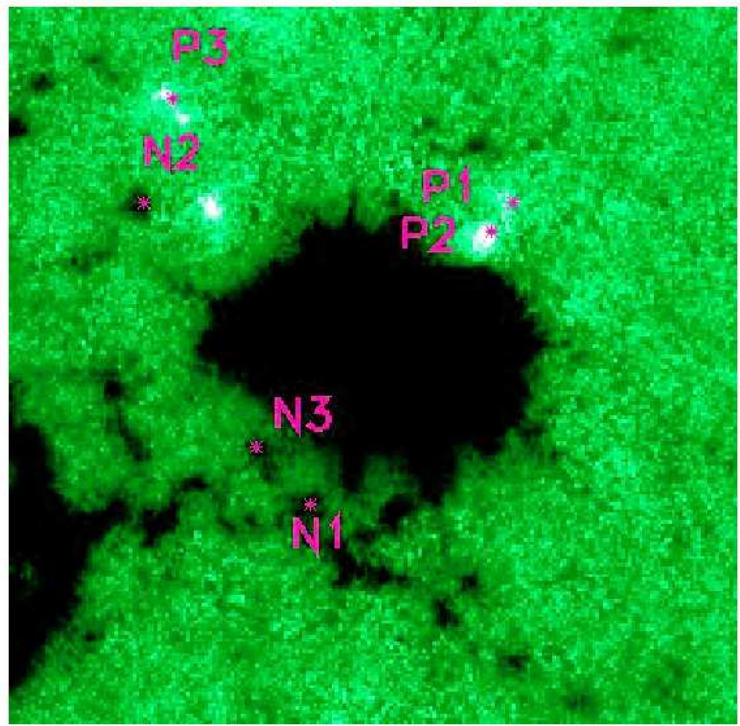}

\caption{Still frame from the Bmovie.avi showing the evolution of the magnetic flux around the pore. All the six MMFs investigated, labeled P or N according to their polarity, are present in this excerpt. 
}
}
\label{fig2_appendix}
\end{figure*}

\textbf{The movie \textit{Bmovie.gif} shows the evolution of the LOS magnetic flux derived from COG in an area surrounding the pore. A subsonic filter was applied to the sequence.  To reduce aliasing effects, the FOV was apodized and reduced to 33.3"$\times$33.3". The magnetic flux is clipped at -200 and 100 G to highlight the formation and evolution of the investigated MMFs. A still frame from the movie, acquired at 17:21:29UT, is shown in Fig.\ref{fig2_appendix}. } 

\end{appendix}
\end{document}